\documentclass[manuscript]{aastex}

\shorttitle{Schmidt-Kennicutt law for the Milky Way}
\shortauthors{Fuchs et al.}

\begin{document}


\title{A Schmidt-Kennicutt law for star formation in the \\ Milky Way disk}


\author{B. Fuchs and H. Jahrei{\ss}}
\affil{Astronomisches Rechen-Institut am Zentrum f\"ur Astronomie der
Universit\"at Heidelberg, 69120 Heidelberg, Germany}

\and

\author{Chris Flynn}
\affil{Tuorla Observatory, University of Turku, FI-21500 Piikki\"o, Finland}



\begin{abstract}

We use a new method to trace backwards the star formation history of the Milky
Way disk, using a sample of M dwarfs in the solar neighbourhood which is
representative for the entire solar circle. M stars are used because they show
H$_\alpha$ emission until a particular age which is a well calibrated function
of their absolute magnitudes. This allows us to reconstruct the rate at which
disk stars have been born over about half the disk's lifetime. Our star
formation rate agrees well with those obtained by using other, independent, 
methods and seems to rule out a constant star formation rate.

The principal result of this study is to show that a relation of the
Schmidt-Kennicut type (which relates the star formation rate to the
interstellar gas content of galaxy disks) has pertained in the Milky Way disk
during the last 5 Gyr. The star formation rate we derive from the M dwarfs and
the interstellar gas content of the disk can be inferred as a function of time
from a model of the chemical enrichment of the disk, which is well constrained
by the observations indicating that the metallicity of the Galactic disk has 
remained nearly constant over the timescales involved. We demonstrate that the
star formation rate and gas surface densities over the last 5 Gyrs can be 
accurately described by a Schmidt-Kennicutt law with an index of $\Gamma$ =
1.45$^{+0.22}_{-0.09}$. This is, within statistical uncertainties, the same 
value found for other galaxies.

\end{abstract}




\section{Introduction} \label{sec1}

Tracing the star formation history (SFH) in galactic disks is a crucial aspect
of understanding their evolution. Determination of a detailed SFH is
practically only possible in the disk of the Milky Way, and various methods
have been developed to do so. The most common method, introduced by
\cite{Twa80}, is to model the colour - magnitude diagram (CMD) of stars in the
solar neighbourhood. Model CMDs are computed by adopting an initial mass
function, a metallicity - age relation, sets of isochrones and iterating the
SFH until computed CMDs match observations \citep{Her00,Ber01,Ver02,Cig06}.
\cite{RoPi00} used the chromospheric emission intensity in the Ca H and K
absorption lines in the spectra of F and G stars. However, due to stellar
activity cycles, metallicity effects and other complications, chromospheric
emission dating can become very uncertain (see \citealt{Fel01} for a full
discussion).  A third method, described by \cite{Noh90}, is based on the
luminosity function of white dwarfs, which preserves details of the
SFH. \cite{LaLa03} have demonstrated that a very considerable fraction of 
the field star population in the Milky Way disk might have been formed in open
clusters. These dissolve on fairly short timescales due to dynamical effects
\citep{FuFu04} so that, even though only a small percentage of the stellar
population is actually found in open clusters, the SFH of open clusters might
be indicative for the overall SFH of the Galaxy. However, since open clusters
are typically young, this method can be used to study the SFH only over the
last Gyr of the evolution of the Galactic disk. Despite differences in the
details, the broad result of these studies is that the star formation rate
(SFR) has been declining slowly for the last 5 Gyr, prior to which it was
rising slowly, and has been punctuated by bursts of star formation from time to
time.

In this study we analyze a sample of M dwarfs, whose spectra show H$_\alpha$
emission (dMe), which we have drawn from the Fourth Edition of the Catalogue of
Nearby Stars (CNS4, \citealt{Jah08}).  Like chromospheric Ca H and
K emission, H$_\alpha$ emission is related to magnetic activity in the stars
which is in turn regulated by their internal constitution, and in particular by
the depth of their convection zones \citep{We07}. \cite{Haw00} have
demonstrated, using M dwarfs in nearby open clusters, that the period in which
the stars show H$_\alpha$ emission is (among other stellar properties) closely
correlated with their absolute magnitudes, and ranges from about
10$^{7.5}$ yrs at absolute $V$-band magnitude M$_V$ = 6 to some 10$^{9.7}$ yrs
at an absolute magnitude of M$_V$ = 13. After these timescales, the H$_\alpha$
emission essentially switches off. We have utilised this effect to reconstruct
the SFH of the Galactic disk in a new, independent manner.

In galaxy evolution studies, the SFR is commonly expressed as a
``Schmidt-Kennicutt'' power law function of the surface density of the
interstellar gas, even though the physical processes leading to this relation
are not yet fully understood (cf.~\citealt{Scha08} for a discussion of the
various concepts). A relation between the SFR and the density of interstellar
gas raised to some power was first suggested by \cite{Schmi59} in order to
explain that the vertical scale height of young stellar populations in the
Galactic disk is smaller compared to that of the interstellar gas. This
relation was later modified by \cite{Ken89,Ken98} who correlated the SFR
observed in other spiral galaxies with the surface densities of their
interstellar gas either in concentric rings around the centers of the galaxies
or averaged over their entire optical disks. The SFRs and gas surface densities
of the galaxies studied by \cite{Ken89,Ken98} are based on present day states
of the disks. Similarly \citet{Bo03} have derived a Schmidt-Kennicutt relation
for the disk of the Milky Way based on its present day gas content. They
correlated the radial density distribution of the interstellar gas in the
Galactic disk with the radial variation of the disk's present day SFR, finding
a relation which is steeper than the original Schmidt-Kennicutt relation for
other galaxies. What we address here is whether the Schmidt-Kennicutt law also
holds true in the disk's evolution in the past, particularly over the last 5
Gyrs or so. The M dwarfs permit us to track the SFH, and we use the chemical
enrichment history of the Galactic disk to reconstruct its gas surface density
over time.

Our paper is organized as follows: In the next section we describe our data.
In section 3 we derive the SFH and in section 4 we try to establish a
Schmidt-Kennicutt law for earlier epochs of Galactic evolution.  The
conclusions are summarized in the last section.

\section{Data} \label{sec2}

\begin{figure}
\plotone{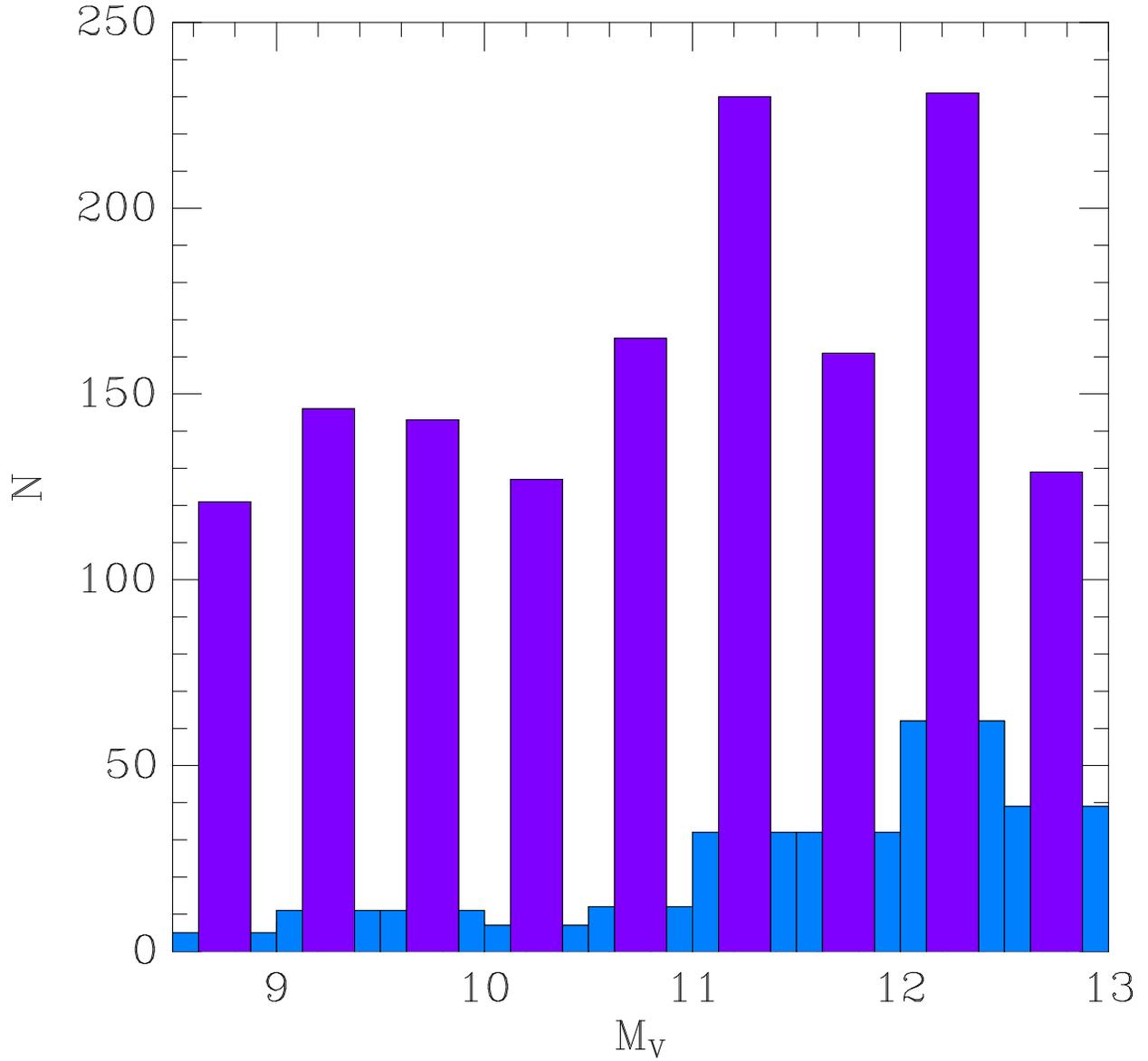}
\caption{Star counts of M dwarfs in the Catalogue of Nearby Stars 
(\citealt{Gli}, \citealt{Jah08}) as function 
of absolute magnitude. The light blue histogram shows the number of 210 M 
dwarfs with H$_\alpha$ emission and the dark blue histogram shows the 
distribution of all 1453 M dwarfs in the magnitude range $M_V$ = 8.75 to 
12.75 mag, respectively.}
\label{fig1}
\end{figure}

Our basic sample consists of 1453 M dwarfs extracted from the CNS4 (4th Catalog
of Nearby Stars) which are within 25 pc from the Sun. 210 of them are
classified as dMe.  The emission strength data come from the Palomar/MSU
Nearby-Star Spectroscopic Survey (\citealt{Rei95}, \citealt{Haw96}). This
survey was based on an earlier version of the Catalogue of Nearby Stars
\citep{Gli}. The dMe are binned into half-magnitude intervals of absolute
magnitudes from M$_V$ = 8.75 to M$_V$ = 12.75 mag and are shown in
Fig.~\ref{fig1} together with star counts in the CNS4 of all M dwarfs of the
same absolute magnitudes. The latter were restricted to that sample on which
the Palomar/MSU Nearby-Star Spectroscopic Survey is based. Since we want to
concentrate here on the thin disk of the Milky Way, we only use stars with
vertical $W$ velocities with $|W| \leq$ 50 km/s \citep{Nor04}.

\begin{figure}
\plotone{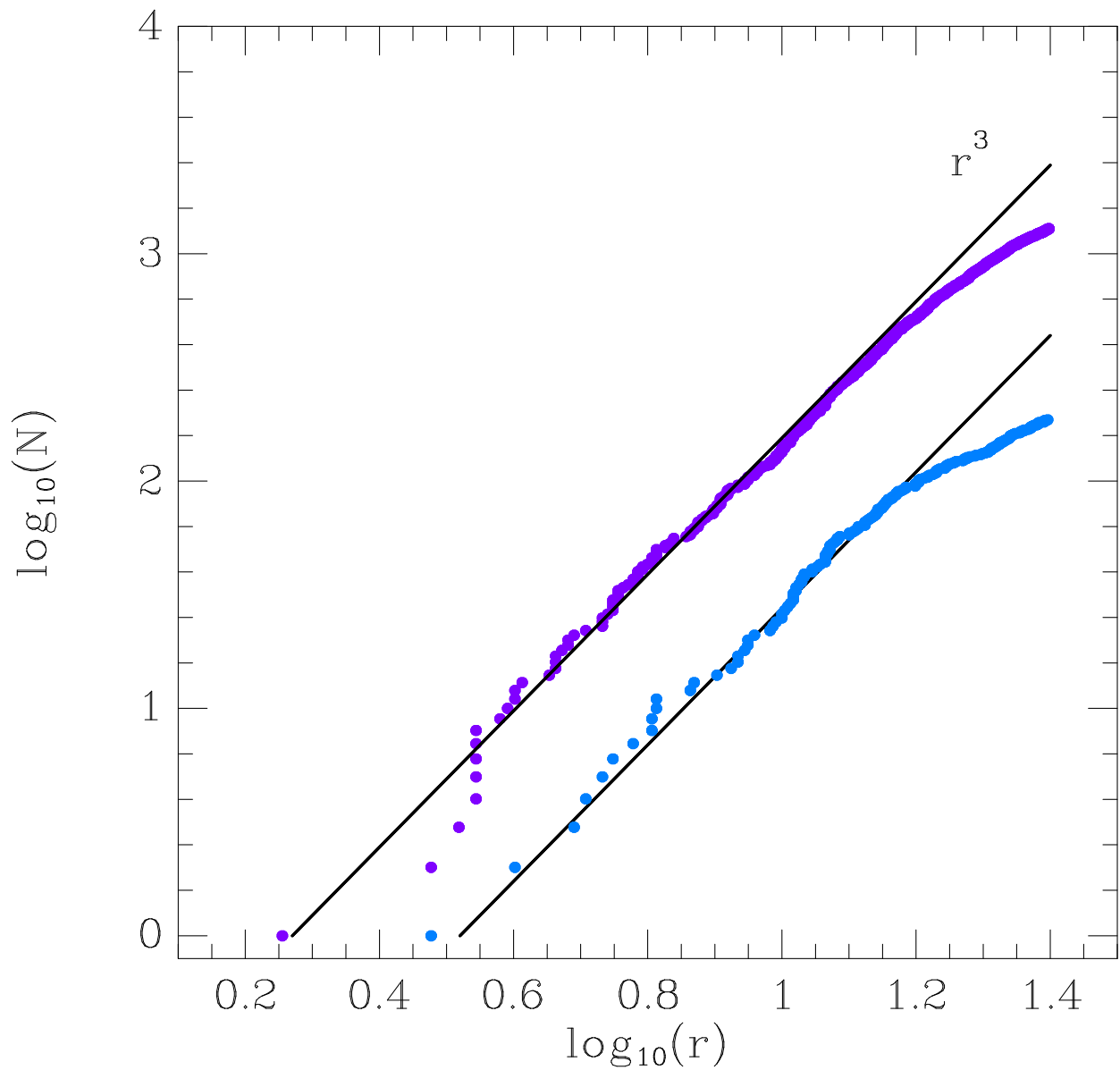}
\caption{Radial cumulative number distribution of M dwarfs with absolute
magnitudes $M_V$ = 8 to 13 mag within the solar neighbourhood drawn as dark blue
dots. The light blue dots show the corresponding radial cumulative number
distribution of M dwarfs with H$_\alpha$ emission. The solid lines with slope 3
indicate the expected radial cumulative number distribution in a homogeneous
sample of stars.}
\label{fig2}
\end{figure}

To assess the completeness of our sample, we have determined the cumulative
number of stars as a function of distance from the Sun. These are shown in
Fig.~\ref{fig2} on double logarithmic scales. In a spatially homogeneous sample
the cumulative number of stars grows in proportion to $r^3$. This is indicated
as solid lines with a slope of 3 in the logarithmic diagram. As can be seen
from Fig.~\ref{fig2} our samples become seriously incomplete beyond 15 pc of
the Sun (this was demonstrated already by \citealt{JaWi97} with a preliminary
version of the CNS4), but this incompleteness has only a small effect on the
method we use, which depends on the relative numbers of stars with and without
H$_\alpha$ emission, and furthermore such stars have almost the same likelihood
of being in the source catalogue. This can be quantified as
follows. Fig.~\ref{fig2} indicates that the dMe are undersampled compared to
the entire sample of M dwarfs. For example, within a radius of 25 pc from the
Sun 53 percent of nearby M dwarfs in the magnitude range, which we consider
here, are catalogued in the CNS4, and 44 percent of the dMe in this range are so
catalogued. To reach a completeness level the same as in the complete sample at
this distance (25 pc) implies that some 43 dMe are missing from the CNS4.

The different sampling rates are almost certainly due to the fact that the M
dwarf sample in the CNS4 is biased towards high proper motion stars. dMe at the
bright end of the luminosity function are very young and have low space
velocities.  Thus it is very plausible that they are undersampled compared to
the entire M dwarf sample. Indeed \cite{Haw96} have shown that the velocity
dispersions of the dMe are significantly smaller than the mean velocity
dispersions of M dwarfs in the CNS4. We discuss possible consequences of the
effect for our results below.

We note that dMe seem frequently to be members of binary or multiple
systems. 63\% of the dMe of our sample are flagged as binaries in the CNS4,
whereas only 28\% of the M dwarfs without H$_\alpha$ emission are so flagged.
However we do not attribute any significance to these numbers, because the
degree of completeness of the identification of binaries in the CNS4 is not
known. In a systematic study, \cite{FiMa92} found that 42\%$\pm$9\% of M dwarfs
in the solar neighbourhood are binaries. Unresolved binarity might occasionally
shift a star in our sample from one half-magnitude bin into a
neighbouring bin, but we expect no systematic effects as a consequence.

Next we calculate the {\it limiting age} $\tau_{limit}$ of the stars in each
magnitude bin, using the calibration of \cite{Haw00}

\begin{equation}
M_V = -16.75 + 3.06\,{\rm log}_{10}(\tau_{limit})\,.
\label{eq1}
\end{equation}

Star formation rates are usually reckoned in terms of surface density per time
interval, whereas our sample is confined to within 25 pc of the Sun, and
samples volume density. To correct our sample from volume to surface density,
we note that in the disk's vertical gravitational potential the vertical scale
height of stars is known to depend linearly on their vertical velocity
dispersion. We have multiplied the numbers of M dwarfs in each magnitude bin by
the $W$ velocity dispersion of stars with the same typical ages to obtain the
surface density of stars of that age, $N'(M_V)=N(M_V)\cdot\sigma_W(\tau)$. For
the latter we have adopted the values given for groups 6d, 6c, 6b, 6a, and
groups 5 - 2 in Table 4 of \cite{JaWi97}.

Our derivation of the SFH relies thus on the frequency of M dwarfs observed in
the solar neighbourhood and corrected to be representative for a cylinder at
the position of the Sun and perpendicular to the Galactic plane. However, the
local volume actually samples stars from much more distant parts of the Milky
Way. At the solar annulus, the orbital period around the Galactic center,
$p=2\pi/\Omega_0$, where $\Omega_0$ denotes the angular velocity of the 
local standard of rest, is $p=2\pi/$ 220 km/s / 8 kpc = 2.3$\cdot$10$^8$
yrs. This means that stars with ages of order 10$^8$ yrs can reach the solar
neighbourhood from the other side of the Galaxy. This effect is enhanced by the
epicyclic motions of the stars. The typical sizes of the epicycles can be
estimated from the velocity dispersions of the stars $\sigma_U$ in the radial
and $\sigma_V$ in the azimuthal directions, respectively, with the relations
\citep{Wie}

\begin{equation}
\sigma_R=\sqrt{\frac{\sigma_U^2}{-4 \Omega_0 B} + 
\frac{\sigma_V^2}{4 B^2}} \quad {\rm and} \quad \sigma_\Phi=
\sqrt{\frac{\Omega_0}{-B}}\cdot\sigma_R\,.
\label{eq1a} 
\end{equation}

$B$ denotes the second Oort constant. If we assume a flat Galactic rotation 
curve, $B=-\frac{1}{2}\Omega_0$, and use again the velocity dispersions given in
Table 4 of \cite{JaWi97}, we find for stars with ages $\tau$ = 0.2 Gyrs 
$\sigma_R$ = 240 pc and $\sigma_\Phi$ = 340 pc, rising to $\sigma_R$ = 1.1 kpc
and $\sigma_\Phi$ = 1.6 kpc for stars with a mean age of $\tau$ = 2.3 Gyrs,
which correspond to our oldest age bin. Thus we sample dMe in a ring around the
Galactic center with a width of about one to two kpc, which we term ``the solar 
circle''.



\section{Star formation history} \label{sec3}

\begin{figure}
\plotone{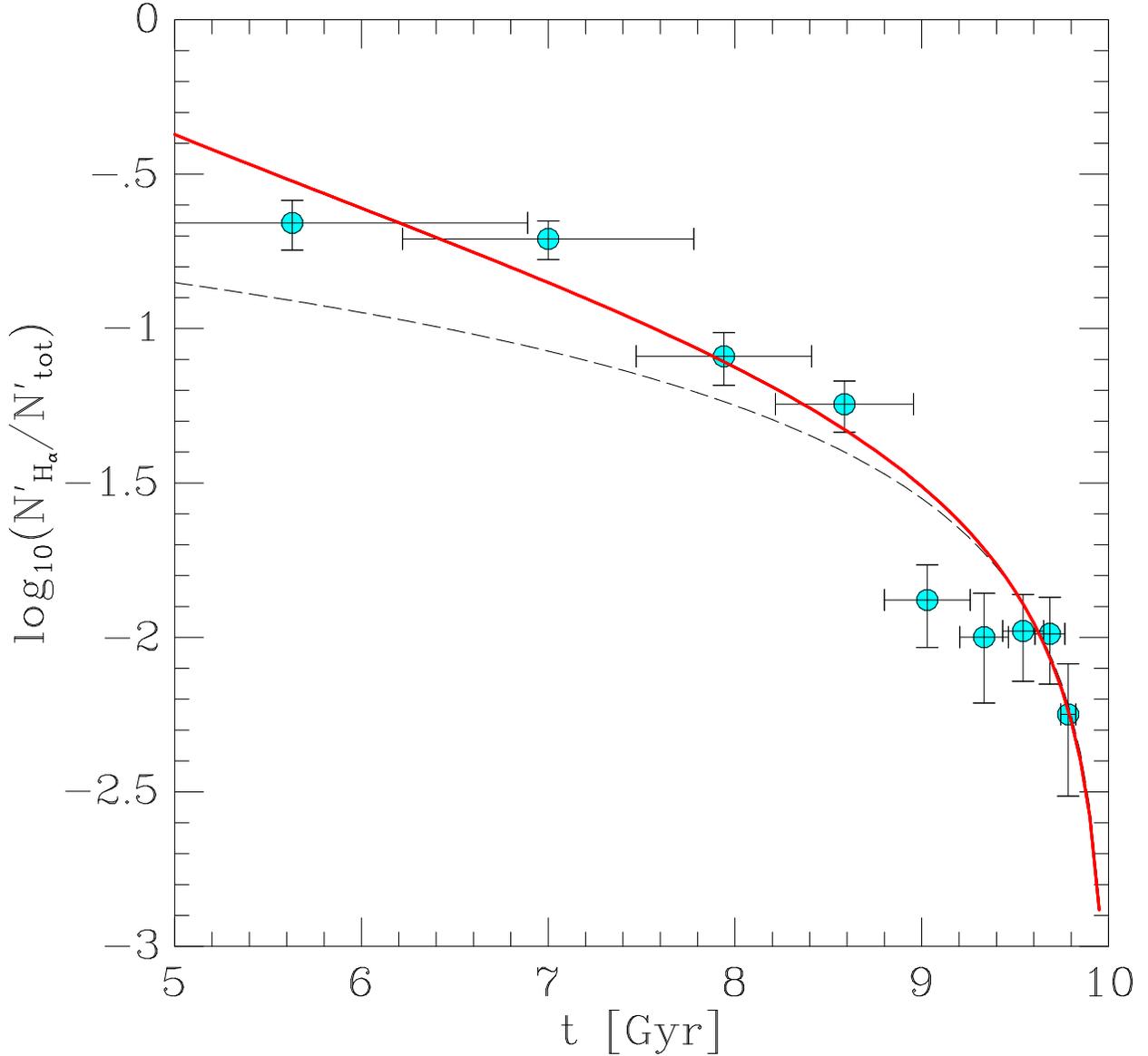}
\caption{The
ratio of the velocity weighted number of M dwarfs with H$_\alpha$ emission
divided by the velocity weighted number of all known M dwarfs at a given age
of the Galactic disk after its formation. We assume a present age of the
Galactic disk of $T$ = 10 Gyrs.
This is a measure of the difference of the surface density of the disk today 
minus the surface density at time $t$, which illustrates the growth of the
stellar disk of the Milky Way. The error bars are explained in the text.
The solid line is a fit of relation (3) with an index of $\beta$ = 2.2 and
a coefficient $\alpha$ = 18.7. The thin dashed line illustrates the growth
of the surface density at a constant star formation rate.}
\label{fig3}
\end{figure}

The surface densities determined in the previous section depend on how far back
the past SFH has been sampled, but are also weighted by the Main Sequence
luminosity function $\Phi_{ms}(M_V)$, i.e.~the relative frequency distribution
of stars as a function of absolute magnitude. Since M dwarfs are so long-lived,
their luminosity function can be safely assumed to be unchanging with time.
As is customary practice in studies of the SFH we assume that the SFH has been 
the same for all M dwarfs in our sample irrespective of mass (they have masses 
in the range 0.7 to 0.2 ${\mathcal{M}}_\odot$, \citealt{Hen}). Below we compare
our results with SFHs determined with other stellar types. The numbers $N'(M_V)$
determined above can be then interpreted as

\begin{eqnarray}
\frac{N'_{H_\alpha}(M_V)}{N'_{tot}(M_V)} & \propto & 
\frac{\Phi_{ms}(M_V)\cdot \int_{T - \tau_{limit}(M_V)}^T 
\dot{\Sigma}_\star(t) dt }
{\Phi_{ms}(M_V)\cdot \int_0^T \dot{\Sigma}_\star(t) dt } \nonumber \\
 & = & \frac{ \Sigma_\star(T)-\Sigma_\star(t)}{\Sigma_\star(T)}\,,
 \label{eq2}
\end{eqnarray}

where $T$ is the present age of the Galactic disk, for which we adopt a value
of 10$^{10}$ yrs, and where $\Sigma_\ast(t)$ and $\dot{\Sigma}_\ast(t)$ denote
the surface density of the stellar disk and the SFR at the epoch
$t=T-\tau_{limit}(M_V)$, respectively. In eq.~\ref{eq2} we have assumed that
the surface density of stellar disk was initially $\Sigma_\star(0)=0$.  Thus
the ratio $N'_{H_\alpha}(M_V)/N'_{tot}(M_V)$ allows us to determine
$\Sigma_\ast(T)-\Sigma_\ast(t)$, which represents the SFH of the disk as a
function of time $t$ relative to the stellar surface density now ($t=T$). This
is shown on logarithmic scale in Fig.~\ref{fig3}. The vertical error bars
reflect only Poisson errors. The horizontal error bars have been estimated 
by the scatter of the calibration data points around the linear 
$M_V$ -- log(age) regression line of \cite{Haw00}. The calibration is based on
5 open clusters in which the authors have determined the limiting $M_V$ of
the dMes. The ages of the clusters are somewhat uncertain. To their oldest
cluster, M67, \cite{Haw00} assign an age of 5.4 Gyrs, whereas other
determinations find 4 $\pm$ 0.5 Gyrs (\citealt{Din95}, \citealt{Mam08}). The
ages of the younger clusters used by \cite{Haw00} like the Hyades or the 
Pleiadesof with ages of 625 and 130 Myrs, respectively, are uncertain on the
order of 10 Myrs \citep{Mam08}.
  
We have fitted to the data in Fig.~\ref{fig3} in the range of disk age
$t$ = 5 to 10 Gyrs a relation of the form

\begin{equation}
\Sigma_\star(t) = \Sigma_{\star0} - \alpha \, t^{-\beta} \,,
\label{eq3}
\end{equation}

and find an index of $\beta$ = 2.2$^{+0.7}_{-0.6}$. The star formation rate 
is then given by

\begin{equation}
\dot{\Sigma}_\star(t) = \alpha \, \beta \,  t^{-(1+\beta)} \,,
\label{eq4}
\end{equation}

a relation which declines with time. As can be seen from Fig.~3 a constant
star formation rate does not fit the data.

As described above, the ratios $N'_{H_\alpha}/N'_{tot}$ have to be corrected
upwards due to the relative undersampling of the dMe. This affects mainly the
younger, bright dMe as they have low space velocities. In our view this
explains the irregular pattern of $N'_{H_\alpha}/N'_{tot}$ ratios at $t$ = 9
and 9.3 Gyrs. Shifting the latter up to the fitting curve in Fig.~\ref{fig3}
would require some 39 additional stars. This is almost precisely the number of
stars we have estimated above needed to be added to the dMe sample so that it
reaches the same level of relative completeness as the entire M dwarf sample. 
As can be seen from Fig.~\ref{fig3} such a correction leads to an even better
fit of eq.~\ref{eq3} to the $N'_{H_\alpha}/N'_{tot}$ data.

The Sun and the stars in its vicinity are at present embedded in the ``Local
Bubble'' of the interstellar medium, a cavity with a diameter of about 200 pc
filled with hot diluted gas. The physical conditions of this gas are such that
one does not expect any stars to be presently forming within it. However, the
Local Bubble is quite young. \cite{Fuc06} have recently redetermined its age
and found that it was formed about 15 Myrs or less ago. The ages of the dMe in
our youngest age bin are of the order of 100 Myrs so that most of them were
born before the Local Bubble was formed. Our results on the SFH are therefore
not affected by the presence of the Local Bubble today.

\begin{figure}
\plotone{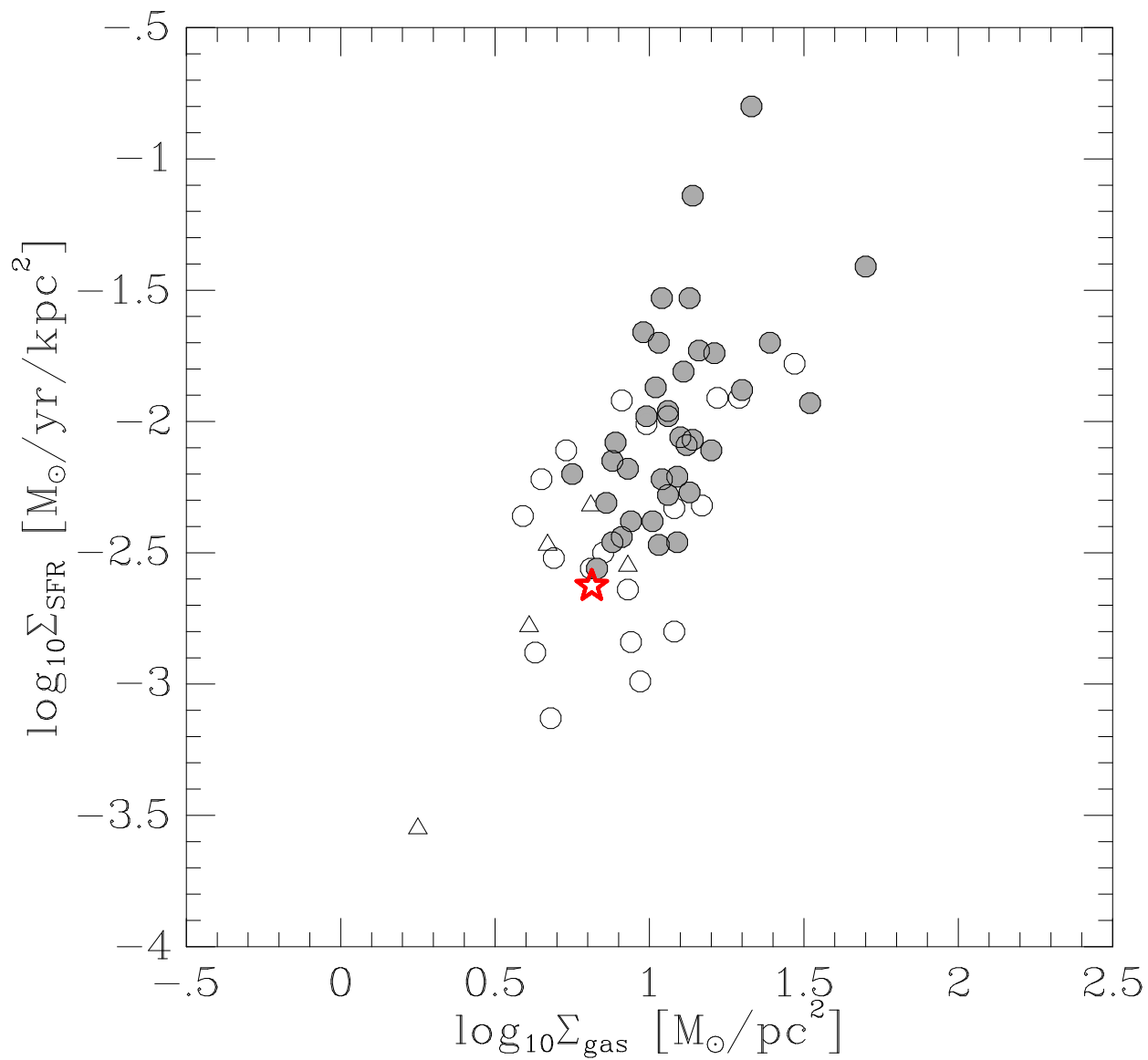}
\caption{The correlation of the star formation rate with the gas surface
density. The symbols are data of 61 normal galaxies reproduced from 
\cite{Ken98}: Sa - Sb galaxies are coded as open triangles, Sb - Sc as open
circles, and Sc - Sd galaxies as filled circles, respectively. The position of
the local Milky Way determined in this work is indicated by a red star.}
\label{fig4}
\end{figure}

Unfortunately we cannot derive absolute rates for the SFH using our
method. However we can estimate its magnitude roughly by observing that,
according to our determination, the surface density of the stellar component of
the Galactic disk's mass grew by 20 percent of today's surface density over the
last 3 Gyrs. \cite{Fly06} find a local surface density of the stellar Galactic
disk of 35.5 ${\mathcal{M}}_\odot$ pc$^{-2}$ which corresponds to a SFR of 2.4
${\mathcal{M}}_\odot$ pc$^{-2}$ Gyr$^{-1}$.  The local surface density of
interstellar atomic and molecular hydrogen is 6.5 ${\mathcal{M}}_\odot$
pc$^{-2}$ \citep{Dam93}. Very interestingly, this combination of parameters
places the Milky Way right on the \cite{Ken98} relation, even though we use
local disk star data, whereas Kennicutt's data are radially averaged over the
whole disk. In Fig.~\ref{fig4} we have reproduced the data of \cite{Ken98} and
indicated the location of the Milky Way in the diagram. As can be seen from
Fig.~\ref{fig4}, the Milky Way falls precisely in the region populated by
similar galaxy types to it, i.e. Sb-Sbc.

\begin{figure}
\plotone{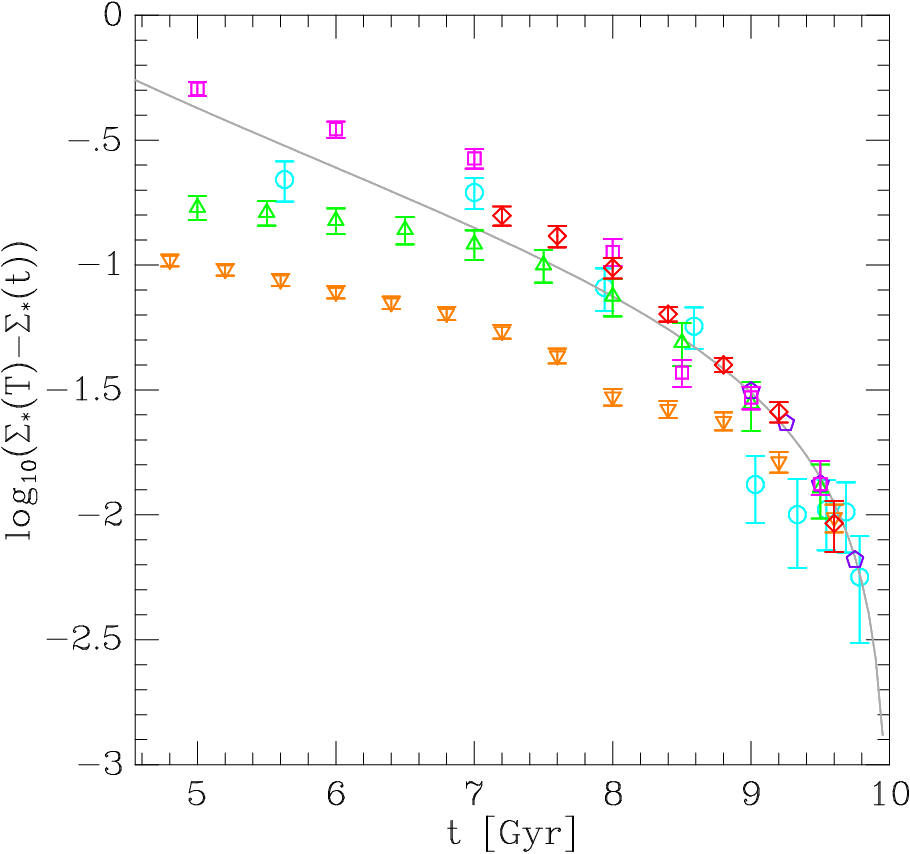}
\caption{The relative star formation history of the Galactic disk 
as determined in this work (circles, same as in Fig.~\ref{fig3}) compared 
with determinations by \cite{Cig06} (squares), \cite{FuFu04} (pentagons),
\cite{Her00} (rhombuses), \cite{RoPi00} (downward triangles), and 
\cite{Ver02} (upward triangles).}
\label{fig5}
\end{figure}

Our determination of the shape of the SFH is consistent with other
determinations of the SFH of the Milky Way disk. We compare our results in
particular with the results of \cite{Cig06}, \cite{FuFu04}, \cite{Her00},
\cite{RoPi00}, and \cite{Ver02}. The results of \cite{Her00}, who modelled the
CMD of stars younger than 3 Gyrs, and \cite{Ver02}, who modelled the CMD of
nearby stars observed with Hipparcos, can be directly compared with our
results, because they refer to column densities. In Fig.~\ref{fig5} we show on
a logarithmic scale $\Sigma_\ast(T)-\Sigma_\ast(t)$ as obtained in this work
versus $\Sigma_\ast(T)-\Sigma_\ast(t)$, which we have constructed by integrating
the SFRs determined by \cite{Her00} and \cite{Ver02}. Short duration star
bursts in the SFR are effectively smoothed away by the integration procedure. 
We show in Fig.~5 also a SFH determined by integrating the SFR derived by 
\cite{FuFu04} from the age distribution of nearby open clusters. Their local 
number densities have been corrected to be representative for column densities
by multiplying them by the vertical velocity dispersions $\sigma(\tau)$ typical
for stars of the same age \citep{JaWi97}. Similarly we have integrated and 
converted the SFR found by \cite{Cig06} modelling the CMD of Hipparcos stars 
(using the data which have been clipped in velocity space). Over the last 3 Gyrs
all SFHs fit excellently together. Given the totally different approaches in
deriving the SFH the good agreement is, in our view, quite remarkable. As
explained above, the deviation of our data 0.7 and 1 Gyr ago from the other SFHs
and the fitting formula \ref{eq3} is in our interpretation due to the proper
motion bias in the CNS4. This affects mainly the bright dMes with short limiting
H$_\alpha$ emission ages. The velocity dispersion of young stars increases with
age by factor of 3 \citep{JaWi97}. Thus our older age bins are expected to be
more complete as they indeed seem to be. Beyond ages of 3 Gyrs the SFH 
determined by \cite{Cig06} is consistent with the SFH derived in this work,
whereas the SFH by \cite{Ver02} falls in that age range for unknown reasons
systematically below the two other determinations. Finally we show in
Fig.~5 also a SFH obtained by integrating the SFR of \cite{RoPi00} which is
based on nearby F and G stars dated with chromospheric emission ages. This SFH
differs significantly from the other SFHs shown there and is consistent with a
constant SFR (cf.~Fig.~\ref{fig3}). With the exception of the 
results by \cite{RoPi00}, the consistent picture of the disk's
SFR is that $\Sigma_\ast(T)-\Sigma_\ast(t)$ rises steeply as a function of
look-back time $T-t$ until about 5 Gyr. We cannot resolve with our data the SFH
at earlier epochs of Galactic evolution. However, the other determinations of
the SFH summarized in \cite{Cig06} indicate a turn over of the SFR to an
apparently much reduced rate at early epochs.

\section{Schmidt-Kennicutt law}

As \cite{Ken98} has pointed out, the data (which we have reproduced in
Fig.~\ref{fig4}) suggest a relation between the SFR and the surface density of
the interstellar gas $\Sigma_g$ of the form

\begin{equation}
\dot{\Sigma}_\star = k \, \Sigma_g^\Gamma
\label{eq5}
\end{equation}

with an index $\Gamma$ = 1.4 $\pm$ 0.15 and a coefficient $k$ =
(2.5$\pm$0.7)$\cdot$10$^{-4}$ $M_\odot$/yr/kpc$^2$.

Let us consider the gas content of the Galactic disk as being regulated by two
processes --- on one hand interstellar gas is converted into stars, but on the
other hand the disk accrets primordial gas,

\begin{equation}
\dot{\Sigma}_g = -\dot{\Sigma}_\star +\dot{\Sigma}_g^{accr}\,.
\label{eq6}
\end{equation}

The present accretion rate of gas, as inferred from infalling high velocity HI
clouds is about 1 ${\mathcal{M}}_\odot$ pc$^{-2}$ Gyr$^{-1}$ \citep{WaWo}, but
the rate is unknown for earlier epochs of Galactic evolution. In order to
constrain the amount of accreted gas we invoke constraints provided by the
chemical enrichment history of the Galactic disk. In a one-zone model the
enrichment or depletion of heavy elements is described by

\begin{equation}
\frac{d}{dt}\left(Z\Sigma_g\right) = -Z \, \dot{\Sigma}_\star +
 y \, \dot{\Sigma}_\star \,,
 \label{eq7}
\end{equation}

where $Z$ denotes the mass fraction of heavy elements. In eq.~\ref{eq7}, we
employed the instantaneous recycling approximation characterized by the yield
$y$. The Geneva - Copenhagen Survey of nearby F and G stars \citep{Nor04} has
shown that the metallicity of the Galactic disk has been, within statistical
uncertainties, constant over most of its evolution (``flat''). In view of
contradictory claims by \cite{Hay06} and \cite{Rei07}, \cite{Hol07} have
thoroughly reanalyzed the Geneva - Copenhagen survey and discussed meticulously
all selection effects. They confirm their earlier conclusion that the
age-metallicity relation is indeed flat. Assuming this to be the case,
equations \ref{eq2} and \ref{eq7} imply then $\dot{\Sigma}_g^{accr}=(y/Z)\,
\dot{\Sigma}_\star$ or

\begin{equation}
\dot{\Sigma}_g  =  - \left(1-\frac{y}{Z}\right)\, \dot{\Sigma}_\star  \,.
 \label{eq8}
\end{equation}

The ratio $y/Z$ has a value of about $y/Z$ = 0.7 \citep{Pag97}, which is
another way of stating that accretion of gas is important for the evolution of
the Galactic disk.

Taking the derivative of eq.~\ref{eq5} we find

\begin{equation}
\frac{ d \dot{\Sigma}_\star}{d\Sigma_g} = \Gamma \, k^\frac{1}{\Gamma} \, 
\dot{\Sigma}_\star ^{\frac{\Gamma - 1}{\Gamma}}\,.
 \label{eq9}
\end{equation}

If we insert the parameterization of $\dot{\Sigma}_\ast$ introduced in
eq.~\ref{eq3}, eqs.~\ref{eq4}, \ref{eq8}, and \ref{eq9} lead to

\begin{eqnarray}
\log{\left(\frac{ d \dot{\Sigma}_\star}{d\Sigma_g}\right)} & = & const_1 +
\frac{1}{\beta}\,\log{\left(\Sigma_{g}-\Sigma_{g0}\right)} \nonumber \\ & = &
const_2 + \frac{1+\beta}{\beta} \,
\frac{\Gamma -1}{\Gamma} \, \log{\left(\Sigma_{g}-\Sigma_{g0}\right)}
\label{eq10}
\end{eqnarray}

with

\begin{displaymath}
\Sigma_{g0} = \Sigma_g(T)-\left( 1-\frac{y}{Z} \right)\cdot \left(
\Sigma_{\star 0} - \Sigma_\star(T) \right) \,.
\end{displaymath}

A comparison of the coefficients shows then that the indices are related by

\begin{equation}
\Gamma = \frac{1+\beta}{\beta} \,.
 \label{eq11}
\end{equation}

In the previous section we determined a value of $\beta$ = 
2.2$^{+0.7}_{-0.6}$, which implies an index of the Schmidt-Kennicutt law of
$\Gamma$ = 1.45$^{+0.22}_{-0.09}$. This fits, within statistical uncertainties,
very well to the index observed by \cite{Ken98} for the present day SFR in 
other galaxies.  Apparently the SFH of the Milky Way disk was regulated by the
same Schmidt-Kennicutt law over the last 5 Gyrs. At earlier epochs the slope 
$\beta$ was probably shallower, which would imply a larger index $\Gamma$ for
the Schmidt-Kennicutt law. 

In deriving eq.~\ref{eq10} we have assumed an accretion rate of primordial gas
that compensated exactly the metal enrichment due to star formation so that the
metallicity stayed constant during the evolution of the Galactic disk, in line
with the observations by \cite{Hol07}. But many models of the chemical
enrichment history of the Galaxy predict a mild chemical enrichment of the
Galactic disk over the last 5 Gyrs (see \citealt{Col08} for a recent
discussion). This is due to less accretion than we assumed. However, our
principal result on the slope of the Schmidt-Kennicutt law does not depend on
the absolute amount of accreted gas. We note that a `closed-box' model with
$\dot{\Sigma}_g^{accr} = 0$, which is known to lead to a relative distribution
of metallicities in the Milky Way disk quite inconsistent with observations
\citep{Pag97}, implies the same relation as we found in
eq.~\ref{eq11}. Cosmological models of gas accretion onto galaxies and of the
ensuing chemical enrichment history predict accretion rates which decline with
time \citep{Col08}, similar to the SFR which we found for the last 5 Gyrs. Thus
a relation of the kind $\dot{\Sigma}_g^{accr} \sim \dot{\Sigma}_\ast$ will
approximately hold and we argue that our result on the slope of the
Schmidt-Kennicutt relation should be rather robust.

\section{Conclusions} \label{sec5}

We have used a new method on a sample of M dwarfs in the solar neighbourhood
showing H$_{\alpha}$ emission in their spectra, to trace backwards the star
formation history (SFH) of the Milky Way disk. The M stars show H$_\alpha$
emission up until a certain age (which is a well calibrated function of their
absolute magnitudes) allowing us to reconstruct the rate at which stars have
been born in the disk at the solar circle over about half its lifetime.

The star formation history of the Galaxy seems to be well resolved over the
last 5 Gyrs using our M dwarf sample. We find that the star formation rate
(SFR) rises continuously when looking backwards to 5 Gyr.  This is in very good
agreement with other determinations of the star formation 
history of the Galactic disk by different methods.

In order to derive a Schmidt-Kennicutt law for the local disk, we have
determined the variation of the surface density of the interstellar gas in the
Galactic disk during the formation and evolution of the disk. The gas content
of the disk is controlled by the competition between the star formation
process, on one hand, and accretion of gas from outside the Galaxy, on the
other hand. Since the latter cannot be observed directly, we had to infer the
amount of accreted gas indirectly. For this purpose we have invoked the
chemical enrichment history of the Galactic disk. Based on the observation that
the metallicity of the Galactic disk has remained rather constant during its
evolution (in particular over the last 5 Gyr) we show that the gas infall rate
is proportional to the SFR.

We have then correlated the SFR with the surface density of the interstellar
gas and found that the correlation followed over the last 5 Gyrs a
Schmidt-Kennicutt law with an index of $\Gamma$ = 1.45$^{+0.22}_{-0.09}$. This
agrees, within statistical uncertainties, with the value of $\Gamma$ = 1.4 found
by \citet{Ken98} for the present day SFR in other galaxies.  It seems that at
epochs of Galactic evolution earlier than that stars formed at a much reduced
star formation rate, which does not follow the Schmidt-Kennicutt law. The
investigation of the reason for this change of the mode of star formation is
beyond the scope of this paper, but we hope to address this question in the
future.



\acknowledgments

We are grateful to Ralf Klessen, who inspired this work. Discussions with Laura
Portinari are gratefully acknowledged. Thanks are also due to the anonymous
referee for valuable suggestions, which helped to improve the paper. This work
was supported in part by the Academy of Finland.




\clearpage



\end{document}